\documentclass{article}
\usepackage{graphicx}
\usepackage{amsmath}
\usepackage{setspace}
\title{Generalized Mass-to-Horizon Entropy and Horizon Thermodynamics}
\author{Smeehan S Shameeem$^1$, Priyesh K V$^1$, Krishna P. B$^2$,  and Titus K Mathew$^3$ \\ $^1$ Department of Physics, St. Paul's College, Kalamassery,  Kochi, 683503, Kerala, India \\$^{2}$Department of Space Science, St. Albert’s College, Ernakulam, 682018, Kerala, India \\$^3$ Department of Physics, Cochin University of Science and Technology, Kochi, 682022, Kerala, India\\
\\ smeehsmeehan@gmail.com; priyeshkv@stpauls.ac.in; krishnapb99@gmail.com; titus@cusat.ac.in}
\date{}

\begin{document}

\maketitle

\begin{abstract}
We investigate the cosmological implications of generalized mass-to-horizon entropy, a two-parameter extension of the standard Bekenstein entropy based on the mass-to-horizon relation. We have derived the modified Friedmann equation using the entropy balance relation through both the Clausius equation and the modified
first law of thermodynamics. The modified Friedmann equation obtained in this way successfully predicts the
transition of the universe into the late-time accelerated expansion phase. We then analyzed the evolution of this
entropy at the horizon of the expanding universe and found that it always increases as the universe expands, and
eventually satisfies the entropy maximization condition at the final state. Further, we analyze the evolution of relative
fluctuations in the energy of the horizon. We found that, the relative fluctuations decreases as the universe expands
and attain a minimum value at the end de Sitter epoch. An interesting to point note is that, the relative fluctuations
depend inversely on the degrees of freedom residing on the horizon, which shows that the horizon behaves like an
ordinary thermodynamic system. Additionally, it was found that, the asymptotic value of the relative fluctuations
depend on the Planck length and the cosmological constant, which indicate that it contains the signatures of the both early and the late universe, especially the dependence on the Planck length implies that, the fluctuations
contain the imprints of the quantum structure of spacetime.


\end{abstract}

\section{Introduction}
\label{Intro}

Recent research in gravity have shown that black hole horizons or cosmological horizons are thermodynamic systems \cite{Padmanabhan:2003gd,Padmanabhan:2009vy}. As a result, they posses both entropy and temperature\cite{PhysRevLett.26.1344,Bekenstein:1973ur,Gibbons:1977mu}. In a seminal paper by Bekenstein, it was showed that the entropy of the horizon is proportional to its surface area\cite{Bekenstein:1973ur}. Meantime, Hawking has shown that the temperature of the horizon is proportional to its surface gravity\cite{PhysRevLett.26.1344}.  Gibbons and Hawking later extended these thermodynamic ideas to cosmological horizons in de Sitter spacetime \cite{Gibbons:1977mu}. Using these ideas, Jacobson\cite{Jacobson:1995ab} have derived the Einstein Field equation of gravity by projecting the Clausius relation, the thermodynamic identity, to a local Rindler horizon by assuming the Bekenstein relation for entropy and Hawking's relation for temperature. This, in turn, implies that the underlying theory of gravity corresponds to Bekenstein relation for horizon entropy is Einstein’s theory of gravity. Although, the Bekenstein entropy relation was a remarkable achievement, it differs from the conventional thermodynamic concept of entropy, where entropy generally scales with the volume of the system rather than with its surface area. This motivates several different approaches in defining the horizon entropy, which are different from the standard Bekenstein considerations.

\par Among the alternative proposals suitable for horizon entropy, the most notable ones are Tsallis-Cirto and Barrow entropies, which aim to account for the quantum and non-extensive effects. The Tsallis-Cirto entropy is given by $S=\gamma A^{\delta}$\cite{Tsallis:2012}, where $\gamma$ is a constant and $\delta$ is the Tsallis parameter. This entropy particularly provides a suitable description for incorporating non-extensive features of the systems with long-range interactions such as gravity.  However, it does not satisfy the additive rule of the conventional thermodynamic entropy. The underlying cosmology corresponding to this entropy was analyzed in \cite{Sheykhi:2018dpn}, where the author derived the modified Friedmann equation from the modified first law of thermodynamics. In reference\cite{Dheepika:2022sio}, the authors performed an almost similar analysis for the Tsallis-Cirto entropy. While Barrow entropy is given by $S= (A/A_0)^{1+\delta/2},$ where $A$ is the horizon area, $A_0$ is the Planck area and $\delta$(here), accounts for possible quantum gravitational modification through a fractal deformation of the horizon structure\cite{Barrow:2020tzx}. The corresponding cosmology to this entropy, in the context of a flat universe, was analyzed in\cite{Sheykhi:2022jqq}, in which the authors derived a modified Friedmann equation and showed that the model could predict the late-time acceleration of the universe at a slightly lower redshift compared to the observed results. 
Yet another form of entropy is the R\'enyi entropy. This entropy ensures the mathematical consistency with improved additivity properties\cite{Renyi:entropy}, since it satisfies the additive property of the standard entropy, due to its logarithmic functional form. It was found that the R\'enyi entropy modifies the Friedmann equations and can lead to non-trivial cosmological evolution, including accelerated expansion and inflationary behavior\cite{Ghaffari:2019mrp}.
A recently introduced entropy is the Kaniadakis entropy, 
$S_{\kappa} = \frac{1}{2\kappa}\left( W^{\kappa}-W^{-\kappa}\right) $ where $\kappa$ is the deformation parameter. This entropy is introduced through the $\kappa$- logarithm, which extends the logarithm in a way consistent with relativistic statistical symmetries\cite{Kaniadakis_2002}. The corresponding modified Friedmann equations contain an effective dark energy component that differs conceptually from the conventional cosmological constant, as discussed in the works\cite{Sheykhi:2023aqa,Lymperis:2021qty}. However, the framework suffers from the absence of a unique theoretical determination of the deformation parameter $\kappa$. Furthermore, the validity of the generalized second law of thermodynamics in such generalized entropic cosmologies depends sensitively on the specific cosmological setup, horizon choice, and parameter ranges are considered. Another generalization of the horizon entropy is the Sharma–Mittal entropy, which provides a two-parameter extension of both the Tsallis-Cirto and R\'enyi entropies. When applied to the apparent horizon within the framework of horizon thermodynamics, it leads to modifications of the standard Friedmann equations. The resulting cosmological equations contain an effective entropic contribution that behaves similarly to a dark energy like component in the context of emergent gravity and horizon thermodynamics. This correction is generally nonlinear in the Hubble parameter due to the underlying non-extensive entropy structure and can for suitable parameter values, lead to late-time accelerated expansion as shown in various Sharma–Mittal cosmology models\cite{Ali:2021yhk,Naeem:2023tcu}. However, the framework introduces additional free parameters, leading to a potential degeneracy in cosmological constraints, and its application remains largely phenomenological without a unique microscopic derivation. Recently, generalized entropy frameworks have received considerable attention in cosmology, where several known non-extensive entropy models can be recovered through suitable choices of the corresponding parameters. The corresponding modified FRW cosmology arising from such generalized entropies and horizon thermodynamics has been extensively investigated in various contexts including holographic cosmology, entropic cosmology, and non-singular bounce scenarios \cite{Nojiri:2022aof,Nojiri:2022dkr,Nojiri:2022nmu,Odintsov:2022qnn,Odintsov:2023vpj,Nojiri:2024zdu}. These studies indicate that generalized entropy provides a consistent framework for studying the thermodynamic aspects of gravitational dynamics and cosmic evolution.

Despite the considerable progress in entropy-based cosmological models, most proposed horizon entropy formulations remain predominantly geometric in nature, since they are generally constructed as functions of the horizon area\cite{Bekenstein:1973ur,Gibbons:1977mu}. Consequently, such approaches do not explicitly incorporate the energy or information content enclosed within the horizon, and the resulting dark energy is essentially geometric rather than arising from the vacuum energy or quantum effects contained within the horizon volume\cite{Padmanabhan:2009vy,Jacobson:1995ab}. This may partly explain why such models are not fully successful in addressing the cosmological constant problem. Interestingly, consistency between Hawking temperature and Clausius relation requires the introduction of a generalized mass-to-horizon relation\cite{Cai:2005ra,Akbar:2006kj}. In this framework, the horizon entropy becomes directly connected to the enclosed energy content, thereby providing a more physically meaningful thermodynamic description of dark energy and gravitational thermodynamics in cosmology.

In the literature, some studies have explored cosmological models based on generalized mass-to-horizon entropy formulations. For example, Basilakos et al \cite{Basilakos:2025wwu} investigated the cosmological implications of generalized mass-to-horizon entropy in the context of the late-time acceleration of the Universe. These authors obtained the modified Friedmann equations by applying the Clausius relation. Through this, they showed that late-time acceleration emerges naturally as an entropic phenomenon associated with the cosmological horizon. A similar method for derving the modified Friedmann equation was presented in \cite{Sheykhi:2025ydd}. Despite these studies, we believe that a comprehensive analysis of the thermodynamic evolution of the horizon in the context of the generalized mass-to-horizon entropy is still lacking. An important point to note is that, in previous works, the modified Friedmann equation was derived using the conventional Clausius relation. However, since mass to horizon entropy contains additional dependence on the horizon radius, it may naturally lead to an extra non equilibrium  entropy production term. This suggests that a more general entropy balance relation, rather than the standard Clausius relation, should be used in deriving the corresponding Friedmann equations\cite{Eling:2006aw}. Accordingly, the first aim of the present work is to derive the Friedmann equations within this modified non equilibrium thermodynamic framework. The second aim is to examine the evolution of horizon entropy itself. We show that, the mass-to-horizon entropy relation, the horizon entropy approaches a maximum value at the final stage of cosmic evolution, similar to the behavior of ordinary thermodynamic systems approaching equilibrium. Following this, we investigate whether this entropy maximization indeed corresponds to a final equilibrium state of the Universe. Furthermore, we analyze the evolution of thermal fluctuations in the horizon energy and examine whether these fluctuations approach a minimum value in the asymptotic late-time epoch.

The paper is organized as follows. In Section 2, we provide a brief overview of the generalized mass-to-horizon entropy. In Section 3, we derive the modified Friedmann equation using the entropy balance relation. Section 4 is devoted to the study of the evolution of the horizon entropy and the conditions required for entropy maximization. In Section 5, we analyze the evolution of the relative  thermal fluctuations of the horizon energy and discuss its significance. In Section 6, we present our conclusions.

\section{Overview of Generalized Mass-to-Horizon entropy}
\label{sec.2}

The standard Bekenstein entropy provides a purely geometric characterization of the horizon entropy, since it is proportional to the area oonly\cite{Bekenstein:1973ur,Hawking:1975vcx}. However, the Clausius relation implicitly involves an exchange of energy across the boundary, which consequently affects the content of the system enclosed by the horizon. This motivates the construction of an entropy that incorporates the energy content within the horizon\cite{Padmanabhan:2009vy,Verlinde:2016toy}. As we mentioned in the introduction, the mass associated with the horizon can be interpreted as a measure of the total energy enclosed within it. Although the detailed internal structure of the system cannot be uniquely determined, the mass encodes the contributions of all forms of energy  present within it. Therefore, constructing a generalized entropy relation based on the mass-horizon connection provides an extension of the standard geometric approach to define the horizon entropy. 

From the basic Black hole theory, it is seen that the characteristic length scale of the black hole horizon is determined by the mass of the system. In the case of a Schwarzschild black hole, the horizon radius is given by $r_s = \frac{2 GM}{c^2}$, which gives the direct proportionality between the mass and the horizon size. This relation indicates that the geometric properties of the horizon are linked to the energy content enclosed within it. Motivated by this, it was proposed that in the context of cosmology the mass enclosed by the horizon may generally scale  non-linearly with horizon length. Therefore, a generalized mass-to-horizon relation was introduced as\cite{Gohar:2023lta,Gohar:2023hnb},
\begin{equation} 
	M =\gamma \frac{c^2}{G} L^n ,
\end{equation}
where $M$ is the mass enclosed within the horizon, $L$ is the cosmological horizon size, $n$ is a non-negative generalized scaling parameter, $c$ is the speed of light, and $\gamma$ is a parameter with dimension  $[L]^{1-n}$, which effectively encodes the underlying physical scale associated with the generalized scaling behavior. For standard Einstein gravity, $n = 1$ whereas for generalized models $n  \neq 1$. This generalized mass-to-horizon relation provides the foundation for constructing modified entropy expressions, enabling a thermodynamic framework that incorporates the energy content of the system in a more fundamental manner. The generalized mass-to-horizon entropy can then be expressed as\cite{Gohar:2023lta,Basilakos:2025wwu}
\begin{equation}\label{eqn:entropy1}
	S_n = \frac{2\gamma n}{n+1} r^{n-1} S_{BH} = C r^{n-1} S_{BH}=\frac{C}{4G} \, r^{n-1} A, 
\end{equation}
where $\displaystyle C={2\gamma n}/({n+1})$, $r$ is the radius of the horizon, $S_{BH}=A/4G$ is the Bekenstein entropy and $A$ is the area of the horizon. As one can see, in the case $\gamma = n = 1 $ one recovers both the standard mass-to-horizon relation, as well as the standard Bekenstein-Hawking entropy \cite{Bekenstein:1973ur,Hawking:1975vcx}.

\section{Friedmann Equations from entropy balance relation}
\label{sec:3}
The evolution of the Universe is conventionally described by the Friedmann equations, which were originally obtained from Einstein’s field equations under the assumptions of homogeneity and isotropy of the Universe, described by the Friedmann–Robertson–Walker (FRW) metric\cite{Friedmann:1924bb,Lemaitre:1927zz}. However, developments in gravitational physics have revealed a profound connection between gravity and thermodynamics. This connection suggests that the equations governing the evolution of the Universe can be derived from the thermodynamics of the horizon by assuming appropriate relations for the horizon temperature and entropy.

In this section, we derive the modified Friedmann equations using two distinct thermodynamic approaches, namely the Clausius relation and the modified first law of thermodynamics, by applying them to the apparent horizon of the expanding Universe. It should be noted that the exact form of the modified Friedmann equations depends strongly on the form of the entropy relation one assumes for the horizon. In the current context, it is the generalized mass-to-horizon entropy. In the literature, the Friedmann equations corresponding to this entropy are derived in\cite{Gohar:2023lta,Basilakos:2025wwu} from both the Clausius relation and the modified first law of thermodynamics. In such derivations, the total change in entropy was taken to be proportional to $\delta Q$, the heat exchanged across the horizon, weighted by the horizon temperature. This is true only if the change in entropy is fully due to the change in the area of the horizon\cite{Eling:2006aw}. However, in more general forms of entropy, which have additional dependencies on the curvature or other equivalent quantities, the total change in entropy contains a thermal contribution proportional to $\delta Q$ and a non-thermal part, which arises due to the variation of the additional curvature dependence of the horizon entropy. Unlike the earlier works in the literature, we have extracted the thermal part from the total variation of the generalized mass-to-horizon entropy, which is then used to obtain the Friedmann equations from the Clausius relation and the modified first law of thermodynamics.

\subsection{Friedmann Equations from Horizon Thermodynamics via the Clausius relation}
\label{3.1}

In an expanding Universe, there is a relation between the horizon entropy and the flux of energy across the horizon. For a horizon at a temperature $T$, the variation of the entropy is related to the variation of the mean energy crossing the horizon by $\delta S = \delta \left<E\right>/T,$ where $\left<E \right>$ is the average energy of the horizon. When this energy transfer is interpreted as heat0, this relation becomes identical with the Clausius relation\cite{Eling:2006aw},  
\begin{equation}
	dS = \frac{\delta Q}{T}.   
\end{equation} 
The energy that flows across the horizon can be considered as heat for any observer near the horizon.

In the context of the Bekenstein relation for the horizon entropy, for which Einstein's field equations represent the underlying gravity, the entropy change, $dS$, is proportional to the change in the area of the horizon. Taking into account of the applicability of the Clausius relation, one can then conclude that a change in the area of the apparent horizon is associated with a heat transfer across the horizon.

In the case of the generalized mass-to-horizon entropy, the horizon entropy is given by the relation (\ref{eqn:entropy1}), i.e., $S=C r^{n-1} S_{BH}$. On varying this, we find,
\begin{equation}\label{eqn:Stot1}
	dS = \frac{C}{4G} r^{n-1} \delta A + \frac{C}{4G} \delta (r^{n-1}) A .
\end{equation}
The first term on the right-hand side of the above equation represents the change in the entropy caused by the change in the horizon area, which, in turn, can be related to the heat flow across the horizon. The second term, which represents entropy variation at constant area, cannot be interpreted as arising from heat flow, but instead corresponds to a non-equilibrium entropy production process occurring in the system. In such a case, we have $dS > \delta Q/T$, so that the true Clausius relation may not hold. This motivates revising the original Clausius relation into a modified one by including an additional non-thermal entropy generated internally within the system, in addition to the entropy generated by heat flow. We can write the entropy balance relation (modified Clausius relation)\cite{Eling:2006aw} as follows,
\begin{equation}\label{eqn:Sbalance}
	dS = \frac{\delta Q}{T} + d_iS.
\end{equation}
With reference to equation (\ref{eqn:Stot1}), we can identify, $d_iS=\frac{C}{4G} \delta (r^{n-1}) A$ and the first term corresponds to the entropy associated within the heat flow, $\delta Q/T.$ Here the temperature $T,$ is typically identified as the standard Hawking temperature of the horizon, given by\cite{PhysRevD.15.2738}
\begin{equation}\label{eqn:T1}
	T = \frac{1}{2 \pi r}.
\end{equation}
We consider an observer asymptotically near the horizon, for whom the horizon is practically stationary. Here, $r$ is the horizon radius. From equation (\ref{eqn:Sbalance}), we can express the thermal entropy as, 
\begin{equation}
	\frac{\delta Q}{T} = dS - d_iS.
\end{equation}
Following equation (\ref{eqn:Stot1}), the above expression can be rewritten as,
\begin{equation} \label{eqn:delQ}
	\frac{\delta Q}{T} = \frac{2\pi C}{G} r^{n} dr.
\end{equation}
The energy inside the horizon is $E=\rho V$, where $\rho$ is the density and $V$ is the volume. Then, $\delta Q$, which characterizes the energy flux across the horizon, can be obtained using the standard relation\cite{Cai:2005ra},
\begin{equation}\label{eqn:de1}
	\delta Q = A(\rho + p) H r dt,
\end{equation}
where $\rho$ is the density of the cosmic component, $p$ is its pressure and $dt$ is an infinitesimally small interval of time.
Using equations (\ref{eqn:T1}), (\ref{eqn:delQ}) and (\ref{eqn:de1}), the Clausius relation can be expressed as,
\begin{equation}
	4\pi G (\rho + p) dt = \frac{2 \gamma n }{(n+1)}\, r^{n-4} dr.
\end{equation}
In order to recover the first Friedmann equation, the above relation must be combined with the conservation equation for the cosmic fluid. In the present treatment, since we are effectively dealing with heat crossing the horizon, the standard continuity relation is sufficient and is given as\cite{Mukhanov:2005sc},
\begin{equation} \label{eqn:conservation}
	\dot{\rho} + 3H(\rho + p) = 0.
\end{equation}
Substituting $(\rho+p)$ from this equation into the previous equation and integrating, we find
\begin{equation}\label{eqn:FR1}
	\frac{8 \pi G}{3} \rho_m = \frac{1}{(n+1)}\frac{4 \gamma n}{(3-n)} H^{3-n} - \frac{\Lambda}{3},
\end{equation}
where we have fixed the integration constant as $\Lambda/3,$ the cosmological constant.

As indicated at the beginning, in the earlier generalized thermodynamic cosmologies, particularly in the works of \cite{Basilakos:2025wwu,Sheykhi:2025ydd}, the modified Friedmann equations were obtained from the variation of the total entropy, including contributions associated with internal entropy production. In the present work, however, a distinction is made between the equilibrium entropy variation related to the horizon heat flux and the entropy generated through non-equilibrium processes. From the standpoint of equilibrium thermodynamics, only the entropy change associated with the heat exchange across the horizon should enter the Clausius relation, while internally generated entropy production terms should be treated separately. It may be noted that, the above modified Friedmann equations is different in\cite{Basilakos:2025wwu,Sheykhi:2025ydd} . However, difference appears only in the numerical coefficients.

This derivation demonstrates that the modified Friedmann equation can be interpreted as a thermodynamic identity arising from the Clausius relation, when applied to the apparent horizon. It therefore supports the viewpoint that gravitational dynamics may be emergent, originating from the thermodynamic behavior of underlying microscopic degrees of freedom associated with spacetime. Moreover, this framework provides a natural foundation for generalized cosmological models, in which modifications of the entropy lead to corresponding deviations from the standard Friedmann equations. Taking $\gamma=n=1,$ the generalized mass-to-horizon entropy becomes the standard Bekenstein entropy and the above equations reduce to the conventional Friedmann equations of Einstein gravity.

\subsection{Friedmann Equations from Horizon Thermodynamics via modified First law of thermodynamics}
\label{3.2}
In this section, we obtain the modified Friedmann equations by applying the first law of thermodynamics to the apparent horizon, with the generalized mass-to-horizon entropy for the horizon. We adopt the first law of thermodynamics in the form given by\cite{Cai:2005ra},
\begin{equation}\label{eqn:1stlaw}
	dE = TdS + WdV,
\end{equation}
where $W=(\rho-p)/2,$ the work energy function. Here we consider the entire bulk of the horizon; as a result, the horizon boundary is taken to be dynamic. Consequently, the temperature relation is modified as\cite{Hayward:1998ee}
\begin{equation}
	T = -\frac{1}{2\pi r}\left(1 - \frac{\dot{r}}{2Hr}\right).  
\end{equation}
which is often referred to as Kodama-Hayward relation for a dynamical horizon. This incorporates the correction arising from the time variation of the horizon radius. 

By combining the relation for temperature and differential change in entropy, we will take care of extracting the thermal part from the total variation of entropy, as we did in the previous section. We then obtain
\begin{equation}
	T \frac{dS}{dt}  = -\frac{1}{2\pi r}\left(1 - \frac{\dot{r}}{2Hr}\right) \frac{C}{G} \, 2 \pi \, r^{n}  \, \dot{r}.
\end{equation}
Combining this with the work energy term leads to
\begin{equation}
	\frac{\mathrm{d}E}{\mathrm{d}t} = -\frac{1}{2\pi r}\left(1 - \frac{\dot{r}}{2Hr}\right) \frac{C}{G} \, 2 \pi \, r^{n}  \, \dot{r} + 2\pi r^2  \frac{(\rho - p)}{2}  \dot{r}.
\end{equation}

As in the previous case, we take the total energy inside the horizon as $  E = \rho V = \frac{4\pi}{3} \rho r^3.$ Using this we can find the following
\begin{equation}\label{eqn:17}
	\frac{\mathrm{d}E}{\mathrm{d}t} = \frac{4\pi}{3} r^3 \dot{\rho} + 4\pi r^2 \rho \dot{r}.
\end{equation}
Combining conservation equation (\ref{eqn:conservation}) with the above expressions(\ref{eqn:17}) obtained for the energy flow, one obtains, (after simplification and equating the two expressions for ($\mathrm{d}E/\mathrm{d}t$)),
\begin{equation}\label{eqn:20}
	\frac{4\pi G}{3} \rho = \frac{1}{(n+1)} \frac{2\gamma_n}{3 - n} r^{n - 3} + C,
\end{equation}
where $C$ is an integration constant. Now substitute for the radius of the horizon of the universe as $r= 1/H$ (for a flat universe). We then get the equation for the density of the cosmic components which yields a modified Friedmann equation,
\begin{equation}\label{eqn:FR2}
	\frac{8\pi G}{3} \rho =  \frac{1}{(n+1) }\frac{4\gamma n}{(3 - n)} H^
	{3 - n} - \frac{\Lambda}{3}
\end{equation} 
where $\Lambda$ is the cosmological constant. 
Equation (\ref{eqn:FR2}), the modified Friedmann equation, is found to be identical to the one obtained using the Clausius relation. 

It may be noted that the integration constant has been chosen to be equal to the cosmological constant,  which causes the late-time accelerated epoch. This may, in turn, imply that the term $\rho$ in the modified Friedmann equations represents radiation and matter respectively, at the relevant epochs\cite{Weinberg:1988cp}. The modified Friedmann equations derived above can explain the late-time evolution of the universe, provided the density $\displaystyle \rho$ = $\rho_m$, the matter density. In the extreme late universe, the cosmological constant $\Lambda$ can dominate over $\rho_m$. Hence, the universe can accelerate. At the end stage, corresponds to the de-Sitter  epoch, the cosmological constant $\Lambda$ becomes fully dominant.

\section{Evolution of horizon entropy and its maximization}
\label{sec.4}

To examine the thermodynamic behavior of the system, we analyze the evolution of the first and second derivatives of the generalized mass-to-horizon entropy using the modified Friedmann equation obtained in the previous section. It is well known that any thermodynamic system must evolve toward a state of maximum entropy corresponding to its equilibrium state. Such an evolution must satisfy two important conditions\cite{Pavon:2012qn,Krishna:2017vmw},
\begin{equation}
	\dot S \geq 0, \quad \quad \ddot S < 0,
\end{equation}
where the dots represent derivatives with respect to cosmic time. The first condition implies that the entropy always increases throughout the evolution of the system. The second condition implies that the entropy approaches a maximum value asymptotically, corresponding to its equilibrium state of the system\cite{Krishna:2017vmw}. We need to check the validity of these two conditions for generalized mass-to-horizon entropy in the context of the expanding universe.

The rate of change in entropy with respect to the cosmic time can be obtained by differentiating  the entropy relation given by the equation with respect to cosmic time (\ref{eqn:delQ}). We then have the following equation
\begin{equation}\label{eqn:dotS}
	\dot S = \frac{dS}{dt} = \gamma \frac{4n\pi}{(n+1)G} r^n \dot{r}. 
\end{equation}
Now the evolution of $\dot S$ is determined by the behavior of $\dot r.$ Using the modified Friedmann equations (\ref{eqn:FR2}) and the conservation relation (\ref{eqn:conservation}), the time derivative of the horizon radius is obtained, as
\begin{equation}\label{eqn:dotr}
	\dot{r} = \frac{2\pi G}{\gamma n} H (\rho + p) (n+1) r^{4-n}
\end{equation}
where $\dot{r}$ depends on the nature of the cosmic fluid.\\
Substituting (\ref{eqn:dotr}) into (\ref{eqn:dotS}), we get
\begin{equation}
	\dot{S} = \frac{4n\pi \gamma}{(n+1)G}  r^{n} \left( \frac{2\pi G}{\gamma n} H (\rho + p) (n+1) r^{4-n} \right) = 8\pi^2 r^4 H \left(\rho+p\right).
\end{equation}
This shows that the evolution of the rate of change of the horizon entropy depends on the evolution of the Hubble parameter $H$ and the quantity $(\rho + p).$ Among these, $H>0$ always, as the universe is expanding; although the magnitude of $H$ decrease with time. The other quantity can be expressed, by considering the barotropic equation of state, $p=\omega \rho,$ as $(\rho+p) = (1+\omega) \rho,$ where $\omega$ is the equation of state parameter of the cosmic component. It has to be noted that $\omega=1/3$ for radiation, $\omega=0$ for matter and $\omega=-1$ for the cosmological constant. The above equation will now attain the form,
\begin{equation}\label{dotsino}
	\dot{S} =  8\pi^2 r^4 H \left(1+\omega \right) \rho. 
\end{equation}

This implies that $\dot S \geq 0$, if $\left(1+\omega \right)\geq 0,$ since the other variables, namely the Hubble parameter and the density are always positive. This, in turn, implies that $\dot S \geq 0$, if $\omega \geq -1.$ This condition is very well satisfied during the radiation-dominated and matter-dominated epochs. During the dark energy-dominated epoch, that is, during the late-time accelerated era, this condition implies that dark energy cannot have phantom nature\cite{Izquierdo:1}. In the current model, the dark energy is represented by an effective cosmological constant, which has an equation of state $\omega=-1;$ and hence, the rate of change of the entropy remains non-negative.
\\

We now investigate the nature of the evolution by differentiating once again, the equation (\ref{dotsino})  with respect to cosmic time, 
\begin{equation}\label{eqn:ddotS}
	\ddot{S} = \frac{4n\pi \gamma}{(n+1)G} \left [ n r^{n-1} \dot{r}^2 + r^{n} \ddot{r} \right].
\end{equation} 
The expression for $\ddot{S}$ consists of two distinct contributions, (i) proportional to $r^{n-1} {\dot r}^2$ and (ii) $r^n \ddot r.$ The first term is always positive; hence the condition for maximization must arise from the second term.

We compute the second derivative of the radius of the horizon by differentiating $\dot{r}$ with respect to cosmic time.  Now with the help of the conservation equation $\dot{\rho} = -3H\rho(1+\omega)$, we get
\begin{align}\label{eqn:ddotr}
	\ddot{r}  & = \frac{2\pi G (n+1)}{\gamma n} (1+\omega)
	\left[(4-n) r^{3-n} \dot{r} H \rho 
	+ r^{4-n} \dot{H} \rho \right. \nonumber \\
	& \qquad \left. - 3 r^{4-n} H^2 \rho (1+\omega)\right]
	+ \frac{2\pi G (n+1)}{\gamma n}
	\left[r^{4-n} H \rho \dot{\omega}\right].
\end{align}
Substituting the above equation into (\ref{eqn:ddotS}), and then taking the asymptotic limit $\omega \to -1,$  at which $\dot r \to 0,$ we observe that 
\begin{equation}
	\ddot S \to \frac{2\pi G (n+1)}{\gamma n} \left[ r^{4-n} H \rho\dot{\omega}\right].
\end{equation}

As the universe evolves from the radiation-dominated era to the matter-dominated era, and finally to the dark energy–dominated phase, the equation of state parameter $\omega$ changes from $1/3$ (radiation) to $0$ (matter) and then to $-1$ (dark energy); i.e., $\omega$ steadily decreases as the universe expands. Hence, its rate of change $\dot{\omega}$ is negative.
As a result, the second derivative of the entropy, $\ddot{S}$ is negative in the asymptotic limit. Therefore, in the final stage, when the cosmological constant dominates, we have $\ddot{S} < 0$. This indicates that the entropy attains a maximum value in the late-time evolution of the universe.

Hence, both the conditions $\dot{S} \geq 0$ and $\ddot{S} < 0$ are satisfied, demonstrating that, the cosmological horizon with generalized mass-to-horizon entropy evolves consistently as a conventional thermodynamic system, whose entropy maximizes at the end stage of the evolution. The maximization of entropy indicates that the system approaches a stable equilibrium state. To examine further, we study the evolution of thermal fluctuations in energy and check whether these fluctuations attain a minimum value at final state. This will be discussed in the next section.

\section{Fluctuations of Horizon Energy}
\label{sec.5}

The cosmological horizon behaves as a thermodynamic system characterized by temperature and entropy. The fluctuations in thermodynamic variables, especially energy, are inherent feature of any thermodynamic system. Fluctuations in the horizon energy, considering the Bekenstein-Hawking entropy for the horizon, have already been analyzed in the literature\cite{Namboothiri:2024xem}. When the horizon entropy takes a different form, its effect on energy fluctuations is worth exploring. In the present section, we analyze thermal fluctuations in the horizon energy following the canonical ensemble formalism\cite{Namboothiri:2024xem,Brandenberger:2006vv,Komatsu:2021ncs}.

In the canonical ensemble approach, we treat the horizon as a canonical thermal system. Applying the canonical ensemble to isolated gravitating systems, such as black holes, leads to divergences in the partition function due to their negative heat capacity\cite{Lynden-Bell:1998pzh,Velazquez_2016}. However, in a cosmological setting, since the observer is inside the horizon, the canonical ensemble becomes a useful tool for analyzing its thermal properties. Many results in horizon thermodynamics are based on Gibbs statistical mechanics. Hence, we employ the canonical ensemble at least as a working framework to study the fluctuations of the horizon energy. In this framework, the statistical properties of the system are described using the canonical partition function defined as
\begin{equation}
	Q(\beta) = \sum_r e^{-\beta E_r},
\end{equation}
where, $(\displaystyle \beta = 1/k_B T_H)$ in which $k_B$ is the Boltzmann constant and the summation extends to all accessible energy states $E_r$ of the system.
The total energy of the thermodynamic system is equivalent to the average over the ensemble $\langle E \rangle$, which is expressed as
\begin{equation}
	E = \langle E \rangle = \frac{\sum_r E_r e^{-\beta E_r}}{\sum_r e^{-\beta E_r}} = -\frac{\partial \ln Q}{\partial \beta}.
\end{equation}
Therefore, the thermal fluctuation in energy $\Delta E,$ is, by definition, the average deviation of the energy from the mean energy,  and it can be logically expressed as\cite{Gibbons:1977mu},
\begin{equation} \label{eqn:delta E}
	\langle (\Delta E)^2 \rangle = \langle E^2 \rangle - \langle E \rangle ^2 = \frac{\partial^2 \log Q}{\partial \beta^2}= -\frac{\partial \langle E \rangle}{\partial \beta}
\end{equation}
where
\begin{equation}
	\langle E^2 \rangle = \frac{1}{Q} \frac{\partial^2 Q}{\partial \beta^2}
\end{equation}
is the average of the square of the energy of the system. The above results can now be applied to the horizon in order to find the fluctuation in the energy of the cosmological horizon.

The energy of the horizon is defined in relation to its temperature $T_H$ and entropy $S_H$ as \cite{Padmanabhan:2012bs},

\begin{equation}\label{eqn:E}
	E = 2 T_H S_H,
\end{equation}
where $T_H$ is the temperature of the horizon adopted from equation (\ref{eqn:T1})\cite{Gibbons:1977mu,Muhsinath:2022cij} and the generalized mass-to-horizon entropy is taken to be $S_H$ from equation(\ref{eqn:delQ}). The above equation can now be expressed as,
\begin{equation}\label{eqn:aveE}
	E = \frac{4\gamma n}{(n+1)^2 \, G }\, \ H^{-n}.
\end{equation}
This shows that the horizon energy depends on the Hubble parameter with power equal to the non-extensive parameter $n$. This equation can be considered, in fact, as the ensemble average of the horizon energy, $\left<E\right>.$

The fluctuation in the energy of the horizon can now be obtained as.
\begin{equation}\label{eqn:fluct}
	\langle (\Delta E)^2 \rangle = \frac{2\gamma n^2}{ (n+1)^2\pi } \frac{\hbar}{G} \, H^{1-n},
\end{equation}
It may be noted that $\hbar/G \sim E^2_P,$ the Planck energy. Now, the above equation can be rewritten as,
\begin{equation}\label{eqn:fluct}
	\langle (\Delta E)^2 \rangle = \frac{2\gamma n^2}{ (n+1)^2\pi } E^2_P \, H^{1-n}
\end{equation}
Hence, it can be stated that the energy fluctuation of the cosmological horizon is of the order of Planck energy, with an additional dependence on the expansion factor $H$. But, Planck energy signals the fundamental quantum scale of spacetime. This means that, even at cosmological scales, fluctuations “remember” the quantum structure of spacetime.

Let us now assume a special case, $\gamma=1, n=1,$ at which the generalized mass-to-horizon entropy reduces to the Bekenstein entropy. The fluctuation will then reduce to the simple relation; 
\begin{equation}
	\langle (\Delta E)^2 \rangle = \frac{1}{ 2\pi} \frac{\hbar}{G} = \frac{E_P^2}{2\pi}.
\end{equation}
which is a perfect constant. So for Bekenstein entropy, which corresponds to Einstein's gravity, the fluctuations in the horizon energy are universal and do not depend on the cosmic epoch. Only for generalized entropies, for instance mass-to-horizon entropy in the present case, the fluctuation do have a dependence on the cosmic history.

The general dependence of the fluctuations on the quantum nature of spacetime can be easily traced to the microscopic structure of spacetime.
To understand that, let us now calculate the relative fluctuation, that is, the fluctuations per average energy. For this we take the ratio of the fluctuation given in equation (\ref{eqn:fluct}) and the square of the average energy given in equation (\ref{eqn:aveE}). We thus obtain; 

\begin{equation}\label{eqn:123}
	\frac{\langle (\Delta E )^2\rangle  }{E^2} = \frac{(n+1)^2\, \hbar G \, H ^ {n+1}}{8 \pi \gamma }.
\end{equation}
In this equation, we can identify the Planck length as  $L_P^2=\hbar G $ and rewrite the equation as 
\begin{equation}
	\frac{\langle (\Delta E )^2\rangle  }{E^2} = \frac{(n+1)^2\, L_P^2 \, H ^ {n-1}}{2 \gamma A_H},
\end{equation}
where $A=4\pi/H^2 $ is the standard equation for the horizon area. The intention for the above way of expressing the relative fluctuation is clear once we re-express the above equation as
\begin{equation}\label{eqn:relfl}
	\frac{\langle (\Delta E )^2\rangle  }{E^2} = \frac{(n+1)^2 \, H ^ {\,n-1}}{2 \gamma \, N_{sur}} ,
\end{equation}
where, $\displaystyle N_{sur} = A/L_p^2$ , as defined in the law of emergence, where $L_P^2,$ is the Planck area corresponds to one degree of freedom\cite{Padmanabhan:2012ik,krishna:2019ufv}. For more details, see \cite{T:2022mqr,Padmanabhan:2019art,Krishna:2022zzw}. The importance of the equation(\ref{eqn:relfl}) is that it reveals the dependence of the relative fluctuation on the microscopic structure of the spacetime through its degrees of freedom. It is to be noted that, as the universe expands, the area of the horizon increases, as a result the degrees of freedom increases, consequently the relative fluctuation will decrease. Apart from this, the relative fluctuation depend also on the cosmic history through the Hubble parameter, which also decreases as the universe expands. Hence the overall effect is that the relative fluctuation in the horizon energy decreases with the expansion of the universe, first due to its direct dependence on $H$ and secondly due to its inverse dependence on $N_{sur}.$ The quite interesting fact is that the inverse dependence of the relative fluctuation on the degrees of freedom also reveals that the horizon behaves as a conventional thermodynamic system.

Again, for the particular case of $\gamma = 1$ and $n = 1$, the expression (\ref{eqn:relfl}) reduces to
\begin{equation}\label{2by nsurf}
	\frac{\langle (\Delta E)^2 \rangle}{E^2} = \frac{2}{N_{sur}}.
\end{equation}
The above equation corresponds to Bekenstein entropy, where the underlying theory is Einstein's gravity, at which the relative fluctuation solely depends inversely on the horizon degrees of freedom. 

From equations (\ref{eqn:relfl}) and  (\ref{2by nsurf}), it is evident that 
\begin{equation}
	\frac{\langle \Delta E \rangle}{E} << 1, 
\end{equation}
in the asymptotic limit, which corresponds to the de sitter epoch, during which dark energy or the cosmological constant, will be the dominant cosmic component. On the other hand, in the early stages, $H$ was extremely large. To make it more clear, equation (\ref{eqn:relfl}) can be suitably expressed as $\left<\Delta E\right> \sim L_P^{-1} H^{(1-n)/2},$ while the horizon energy can be expressed as $E \sim L_p^{-2}H^{-n}.$ Hence the relative fluctuation will be of the order of $\left<\Delta E\right> /E \sim L_P H^{(n+1)/2},$ which becomes infinitesimally small in the final de sitter epoch. But the relative energy fluctuations are large in the early epochs due to the large magnitude of $H.$ 

The results on fluctuation and relative fluctuations can be summarized as follows. The fluctuation on the horizon energy, $\Delta E$ happened to be a constant of the order of $\hbar/G$ throughout the evolution of the universe if the underlying gravity is Einstein's theory, which admits the Bekenstein entropy for the horizon. On the other hand, for the generalized mass-to-horizon entropy, for which the underlying gravity is different from Einstein's theory, the relative fluctuations depend on the expansion rate, $H.$

\subsection{Physical Significance of Fluctuations in Horizon Energy}
\label{sec:5.1}

Since the horizon behaves as a thermodynamic system, fluctuations in its energy are inevitable. We now investigate the physical significance of these fluctuations in the horizon energy.Let us rewrite the relation of the relative fluctuation given in equation (\ref{eqn:123}) as,
\begin{equation}\label{eqn:rfinep}
	\frac{\langle (\Delta E) \rangle} {E} = \frac{(n+1)}{2 \sqrt{2\gamma \pi}} L_P  \, H^{(n+1)/2}.
\end{equation}
This shows that the fluctuation per horizon energy strongly depends on the Hubble parameter. According to the Friedmann equation,given by equation (\ref{eqn:FR1}), the evolution of the Hubble parameter is in turn related to the the cosmic components. , The physical significance of the relative fluctuation, consider the asymptotic condition the dominant cosmic component is the cosmological constant $\Lambda$. Under this condition, the Friedmann equation takes the form,
\begin{equation}\label{eqn:H}
	H \sim \left( \frac{(n+1) (3 - n)}{4 \gamma n} \, \frac{\Lambda}{3}  \right)^{\frac{1}{3 - n}}.
\end{equation}
Taking this into account, the asymptotic expression for the relative fluctuation can be written as, 
\begin{equation}
	\frac{\langle (\Delta E) \rangle}{E} \ \sim \left(\frac{(3-n) (n+1)}{4 \gamma n}\right)^{\frac{(n+1)}{2(3-n)}} \, {\frac{(n+1)}{2\sqrt{2 \gamma \pi}}} \, L_P \left(\frac{\Lambda}{3}\right)^{\frac{(n+1)}{2(3-n)}}. 
\end{equation}
Taking, $n$ is very close to 1 \cite{LUCIANO2026100487} and the above equation can be approximated as 
\begin{equation}\label{eqn:flucF}
	\frac{\langle (\Delta E) \rangle}{E} \sim \frac{1}{\gamma} \frac{1}{\sqrt{2 \pi n} } \, L_P \sqrt{\Lambda}.
\end{equation}

Here, $L_P$, the Planck length represents the ultraviolet cut-off and is the shortest length of spacetime, signifying the quantum nature of space. The product $\sqrt{n} \gamma,$ bears the signature of the generalized mass-to-horizon entropy. The quantities $\gamma$ (with dimensions $(length)^{(n-1)}$) and $n$ (dimensionless) are the parameters determining the strength of the generalized mass-to-horizon entropy relation. The cosmological constant $\Lambda$ corresponds to the infrared scale, the largest associated with the cosmological horizon ($\Lambda^{1/2}$). Dependence on the Planck length and the cosmological constant indicate that the fluctuations of the horizon energy are not merely quantum effects, nor large-scale cosmological effects alone, instead, they emerge from an interplay between microscopic spacetime structure and cosmic expansion.

Using the standard values, $L_P \sim 10^{-35}$ m  and $\Lambda \sim 10^{-52} m^{-2}$ \cite{1SupernovaSearchTeam:1997sck,2SupernovaSearchTeam:1998bnz,3SupernovaSearchTeam:1998fmf,4SupernovaSearchTeam:1998cav,5SupernovaCosmologyProject:1996grv,6SupernovaCosmologyProject:1997zqe,7SupernovaCosmologyProject:1998vns},  it turns out that $\Delta E/E \sim 10^{-61}/(\sqrt{n} \gamma ),$  which is very small. This means that relative fluctuations at the end stage implies that the late time universe will  be minimum at equilibrium corresponds to the final stage of evolution.

Further, its from the above equation that, 
the relative fluctuations are suppressed by the Planck length $L_p$. This suggests that the observed smallness of $\Lambda$ may be linked to statistical fluctuations of the horizon degrees of freedom. 

The horizon energy is given by $E=2T_H S_H$. The fluctuations in energy corresponds to the fluctuations in entropy. A minimum value of the fluctuations then implies that the corresponding fluctuations in entropy is also small at end stage, where the entropy is getting maximized.


\section{Conclusion}
\label{sec.6}
The recently proposed generalized mass-to-horizon entropy relation provides a unique connection between the total gravitating mass enclosed within a cosmic horizon and the entropy associated with that horizon. This construction establishes a direct link between the matter–energy content of the Universe and the thermodynamic or informational capacity of spacetime, thereby offering a framework for studying cosmologic dynamics in relation to the matter content of the universe, compared to other forms of horizon entropies.

We first derived the modified Friedmann equation using the entropy balance relation $dS=\delta Q/T + d_iS,$ unlike previous works. This approach takes into account the non-equilibrium entropy produced due to the additional dependence of the generalized mass-to-horizon entropy on the horizon radius, apart from its usual dependence on the horizon area. We applied this method in the framework of both the Clausius relation and the modified first law of thermodynamics.
 
It is found that, the resulting Friedmann equation differs from those reported in earlier works, in the coefficients of the energy density terms. This modified Friedmann equation predicted the transition from the matter-dominated epoch to late-time accelerated epoch dominated by the cosmological constant. 

We then analyzed the evolution of the horizon entropy.
We found that the entropy is non-decreasing, satisfying $\dot{S} \geq 0$, throughout the evolution of the universe, which is in agreement with the generalized second law of thermodynamics. We also examined whether the generalized mass-to-horizon entropy assumed for the horizon, maximizes at the asymptotic stage. It was found that, in the asymptotic regime, dominated by the cosmological constant the condition $\ddot{S} < 0$ holds, indicating that the entropy approaches a maximum value in the end epoch. This confirms that the Universe asymptotically evolves toward a state of thermodynamic equilibrium, at the end de sitter phase. 
 
We next analyzed the fluctuations in the  horizon energy using the canonical ensemble formalism by treating the horizon  as a canonical system. The primary aim of this analysis was to examine whether the relative fluctuations in the horizon energy decreasing and attaining a minimum value at the end epoch. If so, it can be interpreted as an indication of a stable equilibrium at the end state. An additional aim was to understand, how these fluctuations depend on the microscopic degrees of freedom residing in the horizon. Our analysis, first shows that the fluctuations in horizon energy, $\Delta E,$ are found to be proportional to (i) the Planck energy $E_P$ and (ii) $H^{n-1}$ as given in equation (\ref{eqn:fluct}). This  is different from the corresponding results in references\cite{Namboothiri:2024xem,Komatsu:2021ncs}, for the case of standard Bekenstein-Hawking entropy, where the fluctuations in the horizon energy are a constant proportional to the Planck energy. This implies that the additional dependence of the fluctuations on the Hubble parameter, is a characteristic feature of the generalized mass-to-horizon entropy. 

 We then obtained the relative fluctuations in the horizon energy. The relative fluctuations have a strong dependence on: (i) $L_P$ the Planck length and (ii) the Hubble parameter, as, $H^{(n+1)/2}.$ This has an important implication at the end epoch. Asymptotically, in the end de Sitter epoch, the Hubble parameter reduces  $H \to \sqrt{\Lambda/3\gamma n},$ which is approximately equal to the square root of the cosmological constant. 
  
 This implies that the relative fluctuations in horizon energy are not merely quantum effects, due to their dependence on $L_P,$ but carry the signature of the macroscopic cosmological evolution itself, through their dependence on the Hubble parameter.
 For the special case, $\gamma=1, \, n=1,$ corresponding to standard Bekenstein-Hawking entropy, the relative fluctuations reduce to a constant, which is independent of cosmic expansion. This highlights a clear distinction  between the standard Bekenstein-Hawking entropy and the generalized mass-to-horizon entropy descriptions. According to our results, for the standard values, $L_P \sim 10^{-35}$ m  and $\Lambda \sim 10^{-52} m^{-2}$  it gives that $\Delta E/E \sim 10^{-61}/(\sqrt{n} \gamma ).$ The suppression of the relative fluctuations by the Planck length may suggest a possible connection between the smallness of $\Lambda$ and the smallness of the relative energy fluctuations. More interestingly, the minimum value of the fluctuations at the end epoch, indicates that, the equilibrium stage corresponding to the late-time universe can be interpreted as a stable equilibrium.
 
As the horizon energy is related to the horizon entropy through $E = 2T_H S_H,$ the decrease in energy fluctuations, implies a corresponding reduction in entropy fluctuations. In the asymptotic stage, where the entropy attains its maximum value, these fluctuations become minimal, signaling the approach of the universe toward a stable thermodynamic equilibrium state. Since, the number of degrees of freedom scales with entropy as $N_{sur} = 4S_H$, as per equation(\ref{2by nsurf}) a larger entropy leads to smaller fluctuations. 

Finally we explored the dependence of the relative fluctuations on the parameters $n$ and $\gamma$ of the generalized mass-to-horizon entropy.
It is expected that the parameters are close to unity and therefore do not significantly affect the magnitude of the relative fluctuations, but contribute only to a mild re-scaling.  

Overall, the generalized mass-to-horizon entropy framework successfully leads to a modified Friedmann equation
capable of describing cosmic evolution, including the observed late-time acceleration of the universe. The framework
also satisfies the fundamental thermodynamic requirements, namely that the entropy always increases and eventually
reaches a maximum at the final stage of evolution. Further, the evolution of the relative fluctuations in the horizon
energy depends not only on the characteristic length scales of the early and late universe, but also explicitly on
the evolution of the Hubble parameter itself. This is a distinctive feature of the present framework, which clearly
differentiates it from the standard Bekenstein entropy scenario.\\

\noindent {\bf Acknowledgements}  
 We thank, Elsa Teena P V, Vishnu A Pai, and Vishnu S Namboothiri for useful discussions during this work.


\end{document}